\documentclass[12pt]{article}
\usepackage{amssymb,amsmath}
\begin{document}

\begin{titlepage}

                            \begin{center}
                            \vspace*{2cm}
\large\bf{Generalized diffusion and asymptotics induced by Tsallis entropy}\\

                            \vfill

              \normalsize\sf    NIKOS \  KALOGEROPOULOS$ \ ^\dagger$\\

                            \vspace{0.2cm}
                            
 \normalsize\sf Weill Cornell Medical College in Qatar\\
                          Education City,  P.O.  Box 24144\\
                                  Doha, Qatar\\

                            \end{center}

                            \vfill

                     \centerline{\normalsize\bf Abstract}
                     
                                               \vspace{1mm}
                     
\normalsize\rm\setlength{\baselineskip}{18pt} \noindent We formulate and solve the
diffusion equation over a previously studied field $\mathcal{R}$, whose construction was 
motivated by the Tsallis entropy composition property. We compare this solution with the 
 solutions of the diffusion and of the porous medium equations. We comment on the 
asymptotics of such solutions for large values of their spatial and temporal variables. 
We present conclusions for the generalised operations inspired by the Tsallis entropy 
composition and their relations to hyperbolicity.    
  
                             \vfill


                             \vfill

\noindent\rule{8cm}{0.2mm}\\
\noindent \small\rm $^\dagger$  E-mail: \ \  \small\rm nik2011@qatar-med.cornell.edu\\
\end{titlepage}


                            \newpage

\normalsize\rm\setlength{\baselineskip}{18pt}

          \centerline{\large\sc 1. \ Introduction}

                           \vspace{5mm}

The Tsallis entropy [1], [2] is a nonenxtensive entropic form
which has attracted considerable attention over the last quarter century [2]. 
Although the scope of its applications is currently unknown, there is 
a large body of circumstantial evidence pointing toward its applicability in diverse 
directions not only in Physics, but also in the biological and social sciences as well as in 
economics and the humanities. This is one reason why a considerable amount of effort 
has been  dedicated to understanding not only the scope but also the properties and the 
dynamical foundations of the Tsallis entropy.\\  

It is probably  fair to state that, in spite of all such efforts and some progress made, there is still 
a lot left to be understood as far as the dynamical foundations of the Tsallis entropy are 
concerned [2]. One way to advance toward this end has been by using mesoscopics   
following the examples  of the Langevin and the Fokker-Planck equations [3], [4]. 
In this formalism, methods and results of the diffusion equation have traditionally played 
a dominant role.  Lessons learned  are extended in the formulation and 
properties of non-linear diffusions [5], of equations with fractional derivatives [6], [7] and of 
their solutions involving the q-analogues of the ordinary transcendental functions [2], [8] - [15].  
A wide range, although still unknown, of applicability of such solutions is guaranteed by a 
relatively recent generalization of the Central Limit Theorem [16] - [19]. There has been a 
recent resurgence of interest in formulating and solving  such equations [20] - [25] 
motivated by the properties of the Tsallis entropy.\\

Probably the most important distinction between the Tsallis and the BGS entropies is the
different way they deal with independent events. The BGS entropy is additive, whereas the
Tsallis entropy is not, generically at least. In order make the Tsallis entropy manifestly additive   
[26], [27] introduced a generalized addition. Alternative proposals were considered  in [28].
A generalized product distributive with respect to the generalised sums of [26], [27] was 
introduced  in [29], [30]. Metric and measure properties of the resulting field, induced by the Tsallis entropy
when compared to its Boltzmann/Gibbs/Shannon (BGS) counterparts, were subsequently examined in [31] - [34].  
Despite the successes of the structures of [26], [27], [29], [30], one can be left wondering to what extent
such generalisations are universal, or what parts of the composition property of the Tsallis entropy they 
really encode. In other words, are there any properties which are weakly dependent, or even independent, 
on the underlying field, if that field's operations encode the composition property of the Tsallis entropy?\\

We address this question in the present work by combine the approaches of the two previous paragraphs.
In Section 2, we use one of the fields introduced in [28] as an alternative to that of [26], [27, [29], [30] to construct 
the diffusion equation on it. In section 3, we present the full solution to this diffusion equation and we compare it
to the solution of the ordinary diffusion equation and to solutions of the porous medium equation. 
We determine and compare their asymptotic behaviours and comment on the common features of such solutions. 
This in an attempt to determine properties that seem to persist regardless of the   base field on which the respective 
equations are formulated. Such a property appears to be the underlying hyperbolicity which is induced by the composition
of the Tsallis entropy as seen in Section 4, where some general conclusions and speculations are presented.\\

                           \vspace{8mm}


       \centerline{\large\sc 2. \ A diffusion equation and asymptotics over \ $\mathcal{R}$}

                           \vspace{5mm}

The Tsallis entropy [1], [2]  is in reality a single parameter family of functionals \ $S_q$ \ parametrized by  \ $q\in\mathbb{R}$. \  
Consider a system having a Riemannian manifold $M$ as its sample space, endowed with a probability density function \
$\rho : M \rightarrow [0,1]$ \  which is absolutely continuous with respect to the Riemannian volume \ $dvol_M$. \ Most frequently 
$M$ stands for the configuration or the phase space of the system. The Tsallis entropy of \ $\rho$ \ is given by  
\begin{equation} 
     S_q [\rho ] = k_B \cdot \frac{1}{q-1} \left\{ 1 - \int_M  [\rho(x)]^q \ dvol_M \right\}
\end{equation}
where \ $k_B$ \  is Boltzmann's constant that will be set to one from now on, for simplicity. Consider two subsets $A, B \subset M$. \ 
They are independent if the corresponding probabilities \ $\rho_A, \ \rho_B$ \ obey the composition law
\begin{equation} 
    \rho_{A \ast B} = \rho_A \cdot \rho_B 
\end{equation}
where $A\ast B$ indicates the system resulting from the interaction of \ $A$ \ and \ $B$. \ It is immediate to check that for such \ $A$ \ 
and \ $B$ \ the Tsallis entropy has the composition law  
\begin{equation}
    S_q [\rho_{A\ast B}] = S_q[\rho_A] + S_q[\rho_B] + (1-q) \ S_q[\rho_A] \ S_q[\rho_B]
\end{equation}
This lead  to the definitions [26], [27]  of the generalized sum
\begin{equation}
    x \oplus_q y = x + y + (1-q) xy
\end{equation} 
and of the generalized product
\begin{equation}
   x\otimes_q y = \left\{ x^{1-q} + y^{1-q} - 1\right\}^\frac{1}{1-q},  \hspace{10mm}  x\geq 0, \ y \geq 0
\end{equation}
Unfortunately these two operations do not satisfy the usual distributivity property, since
\begin{equation}
   x \otimes_q (y \oplus_q z) \neq (x \otimes_q y) \oplus_q (x \otimes_q z) 
\end{equation}
The question [2] of defining a generalised product \ $\Diamond_q$ \ which is distributive with respect to the generalised sum 
\ $\oplus_q$ \ (4) was answered recently [29], [30] where such a product was provided in closed form in [29] by 
\begin{equation}
   x \Diamond_q \ y = \frac{(2-q)^{\frac{\log [1+(1-q)x] \cdot \log [1+(1-q)y]}{[\log (2-q)]^2}}  -1}{1-q}
\end{equation}
This gave rise [30] to the field deformation \ $\mathbb{R}_q$ \ of \ $\mathbb{R}$ \ whose metric properties were used to reach 
[30] - [34] some model independent results pertaining to the Tsallis entropy.\\

We observe that the generalised sum (4) is similar to the ordinary sum when \ $|x| \sim 1, \ |y| \sim 1$, \ but it is more akin to the 
ordinary multiplication when \ $|x| \gg 1, \ |y| \gg 1$. \ Motivated by this observation and the requirement for distributivity, 
we proposed in [28] a reversal of roles of the  generalised addition and multiplication. We obtained, as a result,  two fields 
indicated by \ $\mathcal{R}_1$ \ and \ $\mathcal{R}_2$ \  in [28]. In the present paper it is sufficient to just use one of them for our arguments, 
let's say \ $\mathcal{R}_1$,  \ which we will call \ $\mathcal{R}$ \ from now on. We will keep the notation of [28] in the sequel, 
so the non-extensive parameter will be indicated by $k$ and the BGS entropy will be recovered in the limit $k\rightarrow 0$.  \ 
Let \ $x,y \in \mathcal{R}$. \ As a reminder, the addition \ $\stackrel{k}{\oplus}$ \  on \ $\mathcal{R}$ \ was
defined [28] by the same law as in (5), namely
\begin{equation}
 x\stackrel{k}{\oplus}y \ = \ (x^k+y^k-1)^\frac{1}{k}
\end{equation}
It turned out [28], that (8)  is associative, commutative, with neutral element \ $1$ \ and the opposite of \ $x$ \ denoted by \
$\stackrel{k}{\ominus}$ \ was found to be
\begin{displaymath}
 \stackrel{k}{\ominus} x \ = \ (2-x^k)^\frac{1}{k}
\end{displaymath}
The subtraction was, as a result, defined [28] by $x\stackrel{k}{\ominus}y = x\stackrel{k}{\oplus}(\stackrel{k}{\ominus}y)$ \ which resulted in
\begin{equation}
 x\stackrel{k}{\ominus}y \ = \ (x^k-y^k+1)^\frac{1}{k}
\end{equation}
The multiplication of \ $\mathcal{R}$ \ was defined [28] by
\begin{equation}
x\stackrel{k}{\otimes}y \ = \ \left\{
\frac{(xy)^k-x^k-y^k+(k+1)}{k}\right\}^\frac{1}{k}
\end{equation}
It turned out [28] to be associative, commutative, with identity element \ $(k+1)^\frac{1}{k}$ \ and inverse element, for \
$x\in\mathbb{R}\setminus\{ 1 \}$ \ and for a generic value of \ $k$ \
\begin{displaymath}
  \stackrel{k}{\oslash}x \ = \ \left\{ 1+\frac{k^2}{x^k-1}
  \right\}^\frac{1}{k}
\end{displaymath}
The division was defined [28] as \ $x\stackrel{k}{\oslash}y \   = \ x\stackrel{k}{\otimes}(\stackrel{k}{\oslash}y)$ \ and turned out to be
\begin{equation}
 x\stackrel{k}{\oslash}y \ = \ \left\{1 +
 k \ \frac{x^k-1}{y^k-1}\right\}^\frac{1}{k}
\end{equation}
An explicit isomorphism \ $w: \mathcal{R} \rightarrow \mathbb{R}$ \ is provided by the k-logarithm \ $\ln_k(x)$ \ [2], namely
\begin{equation}
         w(x) \ = \ \frac{x^k-1}{k}
\end{equation}
It may be one some interest to compare (12) to the isomorphism \ $\tau_q$ \ between \ $\mathbb{R}$ \ and \ $\mathbb{R}_q$ 
which was given in [30] by
\begin{equation}
   \tau_q (x) = \frac{(2-q)^x - 1}{1-q}
\end{equation}
We observe that in the present case the map \ $f$ follows a power-law (polynomial) whereas \ $\tau_q$ \ is exponential.
Hence the corresponding fields \ $\mathcal{R}$ \ and \ $\mathbb{R}_q$ \ are substantially different from each other. This 
difference can be ascribed to the relative reversal of roles of the generalised additions in \ $\mathcal{R}$ \ and \ $\mathbb{R}_q$. \ 
In \ $\mathcal{R}$ \ (8) reduces to the addition of the Banach space \ $l^k$, \ for \ $k>1$, \ whereas in $\mathbb{R}_q$ \  
(4) is more akin to multiplication for large values of \ $x$ \ and \ $y$. \\  

Among  all rational powers of \ $x\in\mathcal{R}$ \ we will only need the square root \ $\sqrt[\tiny{\textcircled{k}}]{x}, \ \ x\geq 1$ \ in the sequel. 
It is defined by demanding
\begin{equation}
 \sqrt[\tiny{\textcircled{k}}]{x} \stackrel{k}{\otimes} \sqrt[\tiny{\textcircled{k}}]{x}  \ = \ x, \ \ \ \ \  1 \leq x \in \mathcal{R}
\end{equation}
By using (8),  (10) we can easily see that \ $\sqrt[\tiny{\textcircled{k}}]{x}$ \ is related to the ordinary square root by
\begin{equation}
 \sqrt[\tiny{\textcircled{k}}]{x} \ = \left\{ 1 + \sqrt{k(x^k-1)} \right\}^\frac{1}{k}
\end{equation}
We defined [28] the derivative of  \ $f:\mathcal{R}\rightarrow\mathcal{R}$ \ as
\begin{displaymath}
 D_{\tiny{\textcircled{k}}} \ = \ \lim_{y\rightarrow x} \ \{ f(y)\stackrel{k}{\ominus} f(x)\} \stackrel{k}{\oslash} \{ y\stackrel{k}{\ominus} x\}
\end{displaymath}
which turned out to be
\begin{equation}
 D_{\tiny{\textcircled{k}}} f(x) \ = \ \left\{ 1 + \frac{1}{x^{k-1}}\frac{d}{dx}[f(x)]^k\right\}^\frac{1}{k}
\end{equation}
The derivative \ $D_{\tiny{\textcircled{k}}}$ \ turned out to be linear with respect to \ $\stackrel{k}{\oplus}, \ \stackrel{k}{\otimes}$ \ 
and also obey Leibniz's rule with respect to these two operations.  Let \ $\mathcal{R}_+ \ = \ \{x\in\mathcal{R}, \ \ x \geq 1 \} $. \ 
The exponential function [28] \ $\exp_{\tiny{\textcircled{k}}} : \mathcal{R} \rightarrow \mathcal{R}_+$ \ was defined as an 
appropriately normalized eigenfunction of \ $D_{\tiny{\textcircled{k}}}$ \ and turned out to be
\begin{equation}
  \exp_{\tiny{\textcircled{k}}} (x) \ = \ \left( 1+ke^\frac{x^k-1}{k}
  \right)^\frac{1}{k}
\end{equation}
The normalization that we used in (17) to specify the integration constant in the eigenvalue equation was \
$\exp_{\tiny{\textcircled{k}}} (1) = (1+k)^\frac{1}{k}$, \ in complete analogy with the ordinary exponential function \ $e^x: \mathbb{R} \rightarrow 
\mathbb{R}_+$. \ The integral \ $\int_{\tiny{\textcircled{k}}}$ \ in \ $\mathcal{R}$ \ was
operationally defined [28] by demanding
\begin{displaymath}
       D_{\tiny{\textcircled{k}}}\int_{\tiny{\textcircled{k}}} f(x)\stackrel{k}{\otimes}d_{\tiny{\textcircled{k}}}x \ = \ f(x)
\end{displaymath}
where \ $d_{\tiny{\textcircled{k}}}x \ = \ \lim_{y\rightarrow x} y\stackrel{k}{\ominus}x$. \ This integral is given by
\begin{equation}
      \int_{\tiny{\textcircled{k}}} f(x)\stackrel{k}{\otimes}d_{\tiny{\textcircled{k}}}x \ = \left\{ 1+\int[(f(x))^k-1]x^{k-1}dx\right\}^\frac{1}{k}
\end{equation}
and, like the derivative \ $D_{\tiny{\textcircled{k}}}$, \ it is linear with respect to \ $\stackrel{k}{\oplus}, \ \stackrel{k}{\otimes}$ \ as it should be.
Using  (16), the partial derivatives \ $\partial_{{\tiny{\textcircled{k}}}t}$ \ of functions on \ $\mathcal{R}$ \ can be defined by
\begin{equation}
 \partial_{{\tiny{\textcircled{k}}}t}f(x,t) \ = \ \left\{ 1+ \frac{1}{t^{k-1}}\frac{\partial}{\partial t}[f(x,t)]^k\right\}^\frac{1}{k}
\end{equation}
and
\begin{equation}
 \partial_{{\tiny{\textcircled{k}}}x}f(x,t) \ = \ \left\{ 1+ \frac{1}{x^{k-1}}\frac{\partial}{\partial x}[f(x,t)]^k\right\}^\frac{1}{k}
\end{equation} \\

With these definitions, the simplest parabolic equation of the diffusion type that can be written on \ $\mathcal{R}\times\mathcal{R}_+$, \  is
\begin{equation}
          \partial_{{\tiny{\textcircled{k}}}t}f(x,t) \ = \ D\stackrel{k}{\otimes} \partial_{{\tiny{\textcircled{k}}}x}\partial_{{\tiny{\textcircled{k}}}x}f(x,t)
\end{equation}
where \ $D > 1$ \ stands for the diffusion constant. To determine its fundamental solution we need to use an appropriate initial condition.
As is done for the ordinary diffusion equation, we will use a unit source in the initial condition and then proceed to the general solution 
by linearity.  The Dirac delta function in \ $\mathcal{R}$ \ is defined, in complete analogy to \ $\mathbb{R}$, \ by demanding
\begin{equation}
 \delta_{\tiny{\textcircled{k}}}(x) \ =  \ 1, \ \ \ 1\neq x\in\mathcal{R} \ \ \ \ \ \mathrm{and} \ \ \ \ \
                 \int_{\tiny{\textcircled{k}}} \delta_{\tiny{\textcircled{k}}} \stackrel{k}{\otimes}d_{\tiny{\textcircled{k}}}x \ = \ (1+k)^\frac{1}{k}
\end{equation}
which, by using (18), reduces to
\begin{equation}
      \{ 1 + \int\limits_{-\infty}^{+\infty} [(\delta_{\tiny{\textcircled{k}}}(x))^k-1]x^{k-1}dx\}^\frac{1}{k} \ = \ (k+1)^\frac{1}{k}
\end{equation}
This can be expressed in terms of the Dirac delta function on \ $\mathbb{R}$ \ by writing
\begin{equation}
       \int\limits_{-\infty}^{+\infty} [(\delta_{\tiny{\textcircled{k}}}(x))^k-1]x^{k-1}dx \ = \ k\int\limits_{-\infty}^{+\infty}\delta(x)dx
\end{equation}
which gives
\begin{equation}
     [(\delta_{\tiny{\textcircled{k}}}(x))^k-1]x^{k-1} \ = \ k\delta(x)
\end{equation}
which eventually results in
\begin{equation}
     [\delta_{\tiny{\textcircled{k}}}(x)] \ = \ \left\{ 1 + k \ \frac{\delta(x)}{x^{k-1}} \right\}^\frac{1}{k}
\end{equation}
To find the fundamental solution of (21), we impose the initial condition
\begin{equation}
        f(x_o, t=1) \ = \ \delta_{\tiny{\textcircled{k}}}(x_o)
\end{equation}
We find, in complete analogy with the solution to the ordinary diffusion equation in \ $\mathbb{R}\times \mathbb{R}$, \  
due to the isomorphism (12)  that the solution \ $f(x,t)$ \ of the initial value problem of (21) and (27) is
\begin{equation}
       f(x,t) \ = \ \stackrel{k}{\oslash} \sqrt[\tiny{\textcircled{k}}]{4\stackrel{k}{\otimes}\pi \stackrel{k}{\otimes}
            D\stackrel{k}{\otimes}t} \ \ \stackrel{k}{\otimes} \ \ \exp_{\tiny{\textcircled{k}}} \left\{ \stackrel{k}{\ominus} [x\stackrel{k}{\otimes}x]
            \stackrel{k}{\oslash} [4\stackrel{k}{\otimes}D\stackrel{k}{\otimes}t] \right\}
\end{equation}
By analogy with the corresponding properties of the fundamental
solution on \ $\mathbb{R}$, \ we find that
\begin{equation}
     \langle x\stackrel{k}{\otimes}x \rangle_k \ = \ 2 \stackrel{k}{\otimes} D \stackrel{k}{\otimes} t
\end{equation}
where
\begin{equation}
    \langle x\stackrel{k}{\otimes}x \rangle_k \ := \ \int_{\tiny{\textcircled{k}}} x\stackrel{k}{\otimes} x \stackrel{k}{\otimes} f(x,t) \stackrel{k}{\otimes}
                     d_{\tiny{\textcircled{k}}}x
\end{equation}
which, upon substitution of (28) in (30), gives
\begin{equation}
  \langle x\stackrel{k}{\otimes}x \rangle_k \ \sim \ t
\end{equation}
for \ $t\rightarrow\infty$. \ If this result is pulled back to \ $\mathbb{R}$ \ by the inverse of \ $w$ \ given in (12), then we see that 
the solution clearly has a diffusive behaviour something which is not unexpected in view of the algebraic identification of \
$\mathcal{R}$ \ and \ $\mathbb{R}$ \ provided by \ $w$. \  It is far more interesting to see how the asymptotic behaviour of 
the solution appears from the viewpoint of $\mathbb{R}$.  It can be easily checked that, up to an unimportant, for present purposes, 
constant
\begin{equation}
      f(x,t) \sim \left\{ 1+k \ \exp \left[ - \frac{x^{2k}-2x^k+1}{((4\stackrel{k}{\otimes} D)^k -1) t^k - (4\stackrel{k}{\otimes}D)^k + 1} \right]
            \right\}^\frac{1}{k}
\end{equation}
which for large \ $x$ \ and large \ $t$ \ simplifies to
\begin{equation}
      f(x,t) \sim \left\{ 1+k \ \exp \left[ - \frac{x^{2k}}{((4\stackrel{k}{\otimes} D)^k -1) t^k } \right] \right\}^\frac{1}{k}
\end{equation}

The asymptotic solution to the initial value problem is clearly a not diffusion. One can observe that the $k$-th power of \ $f(x,t)$ \ behaves 
like stretched exponential as a function of $x$ and $t$  with an additional constant term  providing the asymptotically dominant, 
even if trivial, behaviour.  This dominant behavior of \ $f^k(x,t)$, \  is to zeroth order similar to the asymptotic temporal behaviour of the 
solutions to the heat equation on a compact manifold [35]. The characteristic exponent determining the rate of approach to the steady state 
is \  $k$ \  and the corresponding ``generalized diffusion" coefficient is   
\begin{equation}
   \mathcal{D} \equiv  \frac{1}{4}[(4\stackrel{k}{\otimes} D)^k -1]^\frac{1}{k} \ = \ \frac{1}{4}\left[\frac{(4D)^k - D^k - 4^k +1}{k}\right]^\frac{1}{k}
\end{equation}
It is may be of some interest to observe that \ $\mathcal{D}$ \ explicitly depends on \ $k$ \ a fact that  becomes more relevant particularly 
when systems, and the resulting fields, described by different values of \ $k$, \ interact with each other. 

                             \vspace{8mm}


 \centerline{\large\sc 3. \ General solution and implications}

                           \vspace{5mm}

\noindent In this section we intend to obtain explicitly a formal general solution of (21).   We proceed by using separation of variables. 
It might be also of some interest to use a  variation of the recently introduced $q$-Fourier transform [17], [36] - [39] and its inverse 
to that end. We speak about a formal solution of (21) since we will not specify any particular initial conditions, unlike (27). To be
on safe ground, an analysis on the existence, convergence and regularity properties [40] of such a solution should also
accompany our result. These are aspects, however, that we will not deal with in the present work.\\

\noindent Substituting  (10), (19) and (20) in (21), we get
\begin{equation}
      \left( 1-\frac{1}{k}\right) + \frac{1}{t^{k-1}}\frac{\partial}{\partial t}[f(x,t)]^k \ = \ \left( \frac{D^k}{k}-1\right)\frac{1}{x^{k-1}}\frac{\partial}{\partial x}
              \left\{ \frac{1}{x^{k-1}}\frac{\partial}{\partial x}[f(x,t)]^k\right\}
\end{equation}
whose general solution we are seeking. This non-linear partial differential equation can be easily reduced to a linear one, if we rewrite it as
\begin{equation}
          \frac{1}{t^{k-1}}\frac{\partial}{\partial t}\left\{ [f(x,t)]^k+\frac{k-1}{k^2} \ t^k\right\} \ = \ \frac{(\frac{D^k}{k}-1)}{x^{k-1}}
                            \frac{\partial}{\partial x} \left\{ \frac{1}{x^{k-1}}\frac{\partial}{\partial x}\{ [f(x,t)]^k + \frac{k-1}{k^2} \ t^k \}\right\}
\end{equation}
and set
\begin{equation}
       h(x,t) \ := \ [f(x,t)]^k + \frac{k-1}{k^2} \ t^k
\end{equation}
which results in
\begin{equation}
       \frac{1}{t^{k-1}}\frac{\partial}{\partial t} \ h(x,t) \ = \ \frac{1}{x^{k-1}} \frac{\partial}{\partial x} \left\{
                      \frac{(\frac{D^k}{k}-1)}{x^{k-1}}\frac{\partial}{\partial x} \ h(x,t) \right\}
\end{equation}
This is a parabolic differential equation. If \ $\frac{D^k}{k}-1>0$, \ then (38) has some resemblance to the diffusion equation on a fractal [35], 
with two additional complications: the diffusion constant on this fractal would be time dependent
\begin{equation}
      \frac{( \frac{D^k}{k}-1 )}{x^{k-1}} \ t^{k-1}
\end{equation}
and there is an overall factor \ $\frac{1}{x^{k-1}}$ \ preceding all derivatives in the right hand side of (38). When \ $D^k<k$, \ depending on the
value of \ $k$, \ we may be able to interpret (38) as a time-reversed version of the diffusion equation as is frequently done in cases of dissipative 
equations. We will assume from now on, for concreteness, that \ $D^k>k$. \ To simplify the subsequent equations, we set 
\ $\tilde{D} = \frac{D^k}{k}-1$ in (38). As mentioned in the first paragraph of this Section, to solve (38) we will follow the somewhat unusual path of 
using separation of variables instead of the more widely used  Fourier transforms. 
We do so, in order to sidestep some, yet unresolved, difficulties associated to the uniqueness of the inverse Fourier transform [37] - [39].
Following the standard practice of separation of variables, we assume that
\begin{equation}
        h(x,t) \ = \ \phi_1(x)\phi_2(t)
\end{equation}
Substituting in (38), we find
\begin{equation}
     \frac{1}{t^{k-1}}\frac{1}{\phi_2(t)}\frac{d\phi_2(t)}{dt} \ = \ \mu \ = 
        \ \frac{1}{\phi_1(x)}\frac{1}{x^{k-1}}\frac{d}{dx} \left\{ \frac{\tilde{D}}{x^{k-1}}\frac{d\phi_1(x)}{dx} \right\}
\end{equation}
where \ $\mu\in\mathbb{R}$ \ is a constant whose specific value is determined by the boundary conditions. The time-dependent part of
(41) can be easily integrated and gives
\begin{equation}
        \phi_2(t) \ = \ \phi_2(1) \ e^{\frac{\mu}{k} (t^k-1)}
\end{equation}
This shows that the time dependence of the solution does not grow uncontrollably as \ $t\rightarrow\infty$, \ when \ $\mu>1\in\mathcal{R}$, \ 
regardless of the  value of the non-extensive parameter \ $k$. \ Generally \ $k\in\mathbb{R}$, \ but for most purposes of interest \ 
$k>0\in\mathbb{R}$ \ [2], which corresponds to \ $k>1\in\mathcal{R}$. \ The spatial part of (41) becomes
\begin{equation}
     \frac{d}{dx} \left\{ \frac{1}{x^{k-1}}\frac{d\phi_1(x)}{dx}\right\} -\frac{\mu}{\tilde{D}}x^{k-1}\phi_1(x) \ = \ 0
\end{equation}
We set
\begin{equation}
     u(x) \ = \ \frac{x^{1-k}}{\phi_1(x)}\frac{d\phi_1(x)}{dx}
\end{equation}
in terms of which (43) gives
\begin{equation}
 \frac{du(x)}{dx} + \left\{ [u(x)]^2-\frac{\mu}{\tilde{D}}
 \right\}x^{k-1} \ = \ 0
\end{equation}
This is a special case of the Riccati differential equation. Although it is not known how to solve, in general, the Riccati differential equation, 
(45) is, luckily, separable.\\

Someone could be wondering about any deeper reasons that might justify the appearance of the Riccati equation (45). 
In geometric terms the origin of (45) in the present context can be traced to the following:  First, we observe that (45) may be re-expressed as
\begin{equation}
 \frac{du}{dz} + u^2(z) - \frac{\mu}{\tilde{D}}  \ = \ 0
\end{equation}
in terms of \ $\mathcal{R}\ni x \ \mapsto \ z \equiv  \frac{x^k-1}{k}\in\mathbb{R}$. \ This is the familiar form of a
Riccati equation that is satisfied by the logarithmic derivative of the Jacobian of a  transformation [41], [42]. 
In the present case \ $u(x)$ \ is almost the logarithmic derivative of \  $\phi_1$ \ as seen on (44). The presence of the 
multiplicative extra term \ $x^{1-k}$ \ in (44) can be traced back to the appearance of the term in the definition (16)
which is, in turn, a result of the isomorphism (12). From this viewpoint, the substitution (44) becomes less mysterious.
Similar things can be can be stated about using the $q$-logarithmic map (12) in going from (45) to (46). What we are 
essential doing is looking at the geometry of \ $\mathcal{R} \times \mathcal{R}$ \ from the viewpoint of \ $\mathbb{R} \times \mathbb{R}$. \
The intrinsic geometry of each one of them is Euclidean. However the geometry of \ $\mathcal{R} \times \mathcal{R}$ \  seen from the
viewpoint of \ $\mathbb{R} \times \mathbb{R}$ \ is non-trivial due to the non-linear nature of the isomorphism $w$ \ (12). Hence the constant time 
lines in \ $\mathcal{R} \times \mathcal{R}$ \ do not have as inverse images constant time hypersurfaces in \ $\mathbb{R} \times \mathbb{R}$.   
Consider the projection of a constant time line of \ $\mathcal{R} \times \mathcal{R}$ \ onto a constant time line of \ $\mathbb{R} \times \mathbb{R}$. 
This projection is non-trivial, as was observed above, and such non-triviality is expressed via the Riccati equation (45). 
This also explains why (45) is a Riccati equation with respect to the tangent vector to the constant time line, rather than 
being a differential equation  with respect to time (normal to the constant time line) direction itself as in [41], [42].  \\

We solve first the special case of (45), for which
\begin{equation}
       u^2(x) \ = \ \frac{\mu}{\tilde{D}}
\end{equation}
for which (45) actually becomes linear. Then
\begin{equation}
      f(x,t) \ = \ \left\{ A(\mu) \ e^\frac{x^k-x_o^k}{k} \ e^{\frac{\mu}{k}(t^k-1)} - \frac{k-1}{k} \ t^k \right\}^\frac{1}{k}
\end{equation}
Here \ $A(\mu)$ \ and \ $x_o$ \ are constants to be determined from the initial conditions. Due to the linearity of (45), and the
lack of any additional constraints for \ $\mu$, \ the general solution of (45) is formally given by
\begin{equation}
     F(x,t) \ = \ \int\limits_1^{\infty} f(x,t) \stackrel{k}{\otimes} \ d_{\tiny{\textcircled{k}}} \mu
\end{equation}
In order for \ $F(x,t)$ \ to be interpreted as a probability distribution function it should satisfy
\begin{equation}
     F(x,t) \geq 1, \ \   \forall  \ (x,t)\in \mathcal{R}\times\mathcal{R}_+
\end{equation}
and it should be normalized in \ $\mathcal{R}$, \ namely it should satisfy
\begin{equation}
     \int\limits_{-\infty}^{+\infty} F(x,t) \stackrel{k}{\otimes} d_{\tiny{\textcircled{k}}}x \ = \ (k+1)^\frac{1}{k}
\end{equation}
In case \ $F(x,t)$ \ is not automatically normalized, the normalization can be achieved by setting
\begin{equation}
      p(x,t) \ = \ F(x,t) \stackrel{k}{\oslash} N
\end{equation}
where
\begin{equation}
      N \ = \ \int\limits_{-\infty}^{+\infty} F(x,t) \stackrel{k}{\otimes} d_{\tiny{\textcircled{k}}}x
\end{equation}
In the general case when
\begin{equation}
      u^2(x) \ \neq \frac{\mu}{\tilde{D}}
\end{equation}
then (45) becomes
\begin{equation}
    \frac{du}{u^2(x) - \frac{\mu}{\tilde{D}}} \ = \ - x^{k-1} \ dx
\end{equation}
which upon integration gives
\begin{equation}
    u(x) \ = \ \sqrt{\frac{\mu}{\tilde{D}}} \ \cdot \ \frac{[A(\mu) + \sqrt{\frac{\mu}{\tilde{D}}}] \ e^{\sqrt{\frac{\mu}{\tilde{D}}}
       \cdot \frac{x^k-x_o^k}{k}} \ + \ [A(\mu) - \sqrt{\frac{\mu}{\tilde{D}}}] \ e^{-\sqrt{\frac{\mu}{\tilde{D}}}
       \cdot \frac{x^k-x_o^k}{k}}}{[A(\mu) + \sqrt{\frac{\mu}{\tilde{D}}}] \ e^{\sqrt{\frac{\mu}{\tilde{D}}}
       \cdot \frac{x^k-x_o^k}{k}} \ - \ [A(\mu) - \sqrt{\frac{\mu}{\tilde{D}}}] \ e^{-\sqrt{\frac{\mu}{\tilde{D}}} \cdot \frac{x^k-x_o^k}{k}}}
\end{equation}
Using (44) and integrating once more, we find
\begin{equation}
    \phi_1(x) \ = \ 2 \ B(\mu) \ \sqrt{\frac{\mu}{\tilde{D}}} \ \cdot \ \frac{e^{\sqrt{\frac{\mu}{\tilde{D}}} \ \cdot \
         \frac{x^k-x_o^k}{k}}}{[A(\mu) + \sqrt{\frac{\mu}{\tilde{D}}}] \ - \ [A(\mu) - \sqrt{\frac{\mu}{\tilde{D}}}] \
            e^{-2 \cdot \sqrt{\frac{\mu}{\tilde{D}}} \ \cdot \ \frac{x^k-x_o^k}{k}}}
\end{equation}
where \ $B(\mu)$ \ is another integration constant that is determined by the initial conditions. Substituting (57) and (42) into (37), 
we eventually find
\begin{equation}
     f(x,t) = \left\{ 2 \ \sqrt{\frac{\mu}{\tilde{D}}} \ C(\mu) \ \frac{e^{\sqrt{\frac{\mu}{\tilde{D}}} \ \cdot \
             \frac{x^k-x_o^k}{k}} \ e^{\frac{\mu}{k} (t^k-1)}}{{[A(\mu) + \sqrt{\frac{\mu}{\tilde{D}}}] \  - \ [A(\mu) - \sqrt{\frac{\mu}{\tilde{D}}}] \
                   e^{-2 \cdot \sqrt{\frac{\mu}{\tilde{D}}} \ \cdot \ \frac{x^k-x_o^k}{k}}}} \ - \ \frac{k-1}{k^2} \ t^k \right\}^\frac{1}{k}
\end{equation}
where \ $C(\mu) = B(\mu) \phi_2(1)$. \ The general solution to the diffusion equation (21) is given by substituting (58)
into (49) for a given set of initial conditions. The required normalization is, once more, provided by (52) and (53).\\

We observe that (58) has the form of a stretched exponential reminiscent of solutions of the  porous medium equation 
to which it may worth comparing. A second, related, reason is that its known solutions exhibit  scaling behaviour [43],  [44] which is an important 
property shared by structures described by the Tsallis entropy, such as fractals, for instance. A third reason for singling out and  paying particular 
attention to solutions of the porous medium equation, among all the possible non-linear and fractional generalisations of the diffusion equation,
is because the porous medium equation arises as the gradient flow of the Tsallis entropy functional [45].\\

Foregoing all specifics, which can be found in standard sources [46], the porous medium equation is the non-linear 
second order quasi-parabolic equation given by
\begin{equation}  
                  \frac{\partial g(\vec{x}, t)}{\partial t}  = \nabla^2 [g(\vec{x}, t)]^m,  \hspace{5mm} m \geq 1
\end{equation}
The Zel'dovich-Kompaneets/Barenblatt  (ZKB) solution is a radially symmetric,  self-similar solution having  
the form, in \ $\mathbb{R}^N \times \mathbb{R}_+$ 
\begin{equation}
           g (\vec{x}, t) \sim \left( \frac{c \ t^{2\zeta} - \lambda \  ||\vec{x}||^2}{t} \right)^\frac{1}{m-1}
\end{equation}
where 
\begin{equation}
       \zeta = (Nm - N + 2)^{-1}, \hspace{8mm}  \lambda = \frac{(m-1)\zeta }{2m}
\end{equation}
and \ $c >0$ \  is a constant depending on the total mass 
\begin{equation}
                    M = \int_{\mathbb{R}^N} g(\vec{x}, t) \  d^N \vec{x}
\end{equation} 
The significance of the ZKB solution (60) is considerable if one takes into account [47]  that under initial data obeying certain conditions, 
the solutions of the porous medium equation converge to (60)  for \ $t\rightarrow\infty$.\ We clearly see two common features between 
(48), (58) and (60). The first one is the appearance of the same exponent,  given in two different notations,  in the overall solution. The second 
is the appearance of a time-dependent ``drift" term in both cases. Naturally, there are differences between (48), (58) on the one hand  and 
(60) on the other: one of them is that (48), (58) have the form of s stretched exponential in the spatial distance, whereas (60) has the form of a 
stretched Gaussian with respect to analogous variable. Moreover,  the temporal asymptotic behaviour of (48), (58) is exponential, 
whereas that of (60) is power-law/polynomial. In addition, we observe that it is the logarithm of (48), (58) that exhibits, asymptotically, scaling 
behaviour as \  $x, t \rightarrow \infty$. \ Equivaletly, it is the exponent of (48), (58) that scales nicely as \ $x, t \rightarrow \infty$. \ 
Indeed, under the transformation   
\begin{equation}
         x \ \mapsto \ \sigma x,     \hspace{10mm}      t \ \mapsto \  \sigma t 
\end{equation}
we observe that 
\begin{equation}
          \log f(x, t) \ \ \   \mapsto \ \ \  \frac{\sigma^k}{k^2} \  \log f(x, t)  
\end{equation}
This scaling behaviour can be traced back, once more, to the exponential form of the isomorphism (12) relating \ $\mathbb{R}$ \ and \ 
$\mathcal{R}$. \ For $k>1$ (48) and (58) can also be interpreted  as log-convex measures on \ $\mathbb{R}^N$, \ which indicates that they 
have common features with the Lebesgue and the Gaussian measures in \ $\mathbb{R}^N$ \ such as their concentration properties [48]. 
Measure concentration is particularly important [49], since it can be seen as one important reason why the predictions of Statistical Mechanics 
result in effectively constant values of the experimentally observed quantities, at least in the thermodynamic limit. 
The concentration of a measure which is not a tensor (power) of another one, allows us to 
extend results that stem from the conventional assumption of independence to far weaker inter-dependencies [50]. In this context, it may be worth 
exploring the concentration properties of measures induced by structures reflecting the composition properties of the Tsallis entropy (3). It may also
be with exploring the relation of such measures, if any, to a recent generalization of the Central Limit Theorem [16] - [19] motivated by and 
tailored to the needs of the Tsallis entropy induced viewpoint toward independence and additivity.

                             \vspace{8mm}


\centerline{\large\sc 4. \ Discussion and conclusions}

                           \vspace{5mm}

 We have found formulated a generalized  diffusion equation on the set \ $\mathcal{R}$ \ whose definition, algebraic and differential 
operations were inspired by the non-extensive properties of the Tsallis entropy [28]. We determined its general solution which happened to 
be a stretched exponential, asymptotically, in both the spatial and time variables. We compared the asymptotic form of our solution to that of the 
ZKB solution of the porous medium equation. We did all this in order to get a better understanding of the implications of the Tsallis entropy 
composition property (3) and subsequently of (4) as compared to the the BGS composition of independent events, which is the regular addition.\\

An extension of this work is motivated by the realisation that  there is not really any compelling reason why the temporal and spatial behavior of a 
system should be described by the same value of the deformation parameter \ $k$ \ in (21), at the mesoscopic level. If the evolution of a system is 
described by a Riemannian or Finslerian metric on its phase space, and if such a metric has a symmetry group that reflects the treatment of the 
spatial and temporal directions on the  same footing and with the same parameters, it may be advantageous to preserve these symmetry properties 
in the system's coarse grained/statistical description. It is important to recall that the diffusion equations are inherently non-relativistic, so they 
manifestly break the, potential underlying Lorentz invariance, if such invariance even exists, of the microscopic system. 
For this reason, it may not be unreasonable to  allow for different values of \ $k$ \ in the spatial and temporal directions of mesoscopic 
equations. Such equations may arise more naturally from 
the composition properties of entropic functionals that depend on two parameters [2]. In such cases, an obvious generalization of (21) would be
\begin{equation}
 \partial_{{\tiny{\textcircled{K}}}t}f(x,t) \ = \
D\stackrel{k}{\otimes}
\partial_{{\tiny{\textcircled{k}}}x}\partial_{{\tiny{\textcircled{k}}}x}f(x,t)
\end{equation}
with \ $K \neq k$. \ It might be of some interest to determine the general solution of (65), if this is practically feasible at all,
and ask for its asymptotic behavior as well as its physical significance, if any. Otherwise, we may want to seek
solutions of (65) by using the ``Tsallis entropy-inspired" q-exponential ansatze [2].\\

As a result of the above analysis, we start seeing a few points that may  be more generic and of broader applicability that supersedes the way 
they were reached in this work. When compared to the ordinary diffusion in $\mathbb{R}^N$, the solutions (48), (58) grow much faster in time.
Hence they are more akin to super-diffusion rather than to regular diffusion. If we still wanted to emulate such a process by a diffusion, we would
say that it would have to take place not on \ $\mathbb{R}^N$ \ but on a space that was negatively curved. This since  on such a space any 
quantity would diffuse faster than in \ $\mathbb{R}^N$, \ as should be intuitively obvious [35]. Hence, form the viewpoint of ordinary diffusion, it is as 
if the use of \ $\mathcal{R}$ \ amounts to endowing \ $\mathbb{R}^N$ \ with a negative curvature metric. So, we arrive at a previously reached result 
[31], where the Tsallis entropy composition (3) induces a hyperbolic structure on the underlying (topological) space. It seems therefore that the 
underlying hyperbolicity encoding (3) is somewhat insensitive to the choice of the base field, be it  \ $\mathcal{R}$ \ as in the present case  
or \ $\mathbb{R}_q$ \ as in [31].\\

A far more general application of this diffusion approach with important implications for geometry be can be followed. One could reverse the above 
line of reasoning: since the diffusion properties  ultimately depend only on the Ricci and not on the sectional curvature per se, one could use the 
behaviour  of the Tsallis entropy, and especially its  convexity [45], to define the Ricci curvature for non-Riemannian spaces [51]. 
One such class of non-Riemannian spaces that has attracted considerable  interest relatively recently in Statistical Mechanics, are 
networks/graphs [52] - [55]. A recently defined Ricci curvature [56] for discrete metric 
spaces, such as graphs, provides an analytical tool for potentially unraveling their pertinent properties. This can be particularly useful 
especially in the research direction of networks in which only a very small subset of its results have been analytically obtained [54], [55] and 
illustrates the robustness as well as the wide range of applicability of the Tsallis entropy formalism.\\   
         
                             \vspace{8mm}


\centerline{\large\sc References}

                            \vspace{5mm}

\noindent [1] C. Tsallis, \ \emph{J. Stat. Phys.} {\bf 52}, \ 479 \ (1988)\\
\noindent [2] C. Tsallis, \emph{Introduction to Nonextensive Statistical Mechanics: Approaching a Complex \\
                                 \hspace*{4.5mm} World}, \  \ Springer, \ New York  (2009)\\
\noindent [3] H. Risken, \ \emph{The Fokker-Planck Equation: Methods of Solution and Applications, 
                                   \\ \hspace*{4.5mm} 2nd Ed.}, \  Springer-Verlag, \ Berlin (1989)\\ 
\noindent [4] N.G. Van Kampen, \ \emph{Stochastic Processes in Physics and Chemistry}, \ Elsevier Sc. B.V., \\
                                  \hspace*{4.5mm} Amsterdam (1992)\\
\noindent [5] T.D. Frank, \ \emph{Non-linear Fokker-Planck Equations: Fundamentals and Applications}, \\ 
                                  \hspace*{4.5mm} Springer-Verlag, Berlin (2005)\\                              
\noindent [6] R. Hilfer, \ \emph{Applications of Fractional Calculus in Physics}, \ (Ed.), \ World Scientific, \\ 
                                   \hspace*{4.5mm}   Singapore  (2000).\\
\noindent [7] J. -P. Bouchaud, A. Georges, \ \emph{Phys. Rep.} {\bf 195},  \ 127 \ (1990)\\
\noindent [8] A.R. Plastino, A. Plastino, \ \emph{Physica A} {\bf 222}, \ 347 \ (1995)\\    
\noindent [9] L.C. Malacarne, R.S. Mendes, I.T. Pedron, E.K. Lenzi, \emph{Phys. Rev. E} {\bf 63}, 030101 (2001)\\
\noindent [10] I.T. Pedron, R.S. Mendes, L.C. Malacarne, E.K. Lenzi, \emph{Phys. Rev. E} {\bf 65}, 041108 (2002)\\
\noindent [11] L.C. Malacarne, R.S. Mendes, I.T. Pedron, E.K. Lenzi, \emph{Phys. Rev. E} {\bf 65}, 052101 (2002)\\
\noindent [12] E.K. Lenzi, L.C. Malacarne, R.S. Mendes, I.T. Pedron, \ \emph{Physica A} {\bf 319}, \ 245  \ (2003)\\
\noindent [13] E.K. Lenzi, R.S. Mendes, L.C. Malacarne, L.R. Da Silva, \emph{Physica A} {\bf 342}, \ 16 \ (2004)\\                                
\noindent [14] I.T. Pedron, R.S. Mendes, T.J. Buratta, L.C. Malacarne, E.K. Lenzi, \ \emph{Phys. Rev. E} \\ 
                                 \hspace*{6.5mm} {\bf 72}, \ 031106 \ (2005)\\
\noindent [15] K.S. Fa, \ \emph{Phys. Rev. E} {\bf 72}, 020101 \ (2005)\\
\noindent [16] U. Tirnakli, C. Beck, C. Tsallis, \ \emph{Phys. Rev. E} {\bf 75}, \  040106(R) \ (2007)\\
\noindent [17] S. Umarov, C. Tsallis, S. Steinberg, \  \emph{Milan J. Math.} {\bf 76}, \ 307 \ (2008)\\
\noindent [18] A. Pluchino, A. Rapisarda, C. Tsallis, \ \emph{Physica A} {\bf 387}, \ 3121 \ (2008)\\
\noindent [19] R. Hanel, S. Thurner, C. Tsallis, \ \emph{Eur. Phys. Jour. B} {\bf 72}, \ 263 \ (2009)\\
\noindent [20] J.S. Andrade Jr., G.F.T. da Silva, A.A. Moreira, F.D. Nobre, E.M.F. Curado, \emph{Phys.\\ 
                             \hspace*{6mm} Rev. Lett.}  {\bf 105}, \ 260601 \ (2010)\\
\noindent [21] F.D. Nobre, M.A. Rego-Monteiro, C. Tsallis, \ \emph{Phys. Rev. Lett.} {\bf 106}, \ 140601 \ (2011)\\ 
\noindent [22] J.S. Andrade Jr., G.F.T. da Silva, A.A. Moreira, F.D. Nobre, E.M.F. Curado, \ \emph{Phys.\\
                             \hspace*{6mm} Rev.  Lett.} {\bf 107}, \ 088902 \ (2011)\\
\noindent [23] F.D. Nobre, M.A. Rego-Monteiro, C. Tsallis, \ \emph{Eur. Phys. Lett.} {\bf 97}, \ 41001 \ (2012)\\  
\noindent [24] M.S. Ribeiro, F.D. Nobre, E.M.F. Curado, \ \emph{ Phys. Rev. E} {\bf 85}, \ 021146 \ (2012)\\
\noindent [25] A.R. Plastino, C. Tsallis, \ \emph{Nonlinear Schroedinger Equation in the Presence of Uniform\\
                               \hspace*{6.5mm} Acceleration}, \  {\sf arXiv:1205.6084}\\
\noindent [26] L. Nivanen, A. Le Mehaut\'{e}, Q.A. Wang, \ \emph{Rep. Math. Phys.} {\bf 52}, \ 437 \ (2003)\\
\noindent [27] E.P. Borges, \ \emph{Physica A} {\bf 340}, \ 95 \ (2004)\\
\noindent [28] N. Kalogeropoulos, \ \emph{Physica A} {\bf 356}, \ 408 \ (2005)\\
\noindent [29] T.C. Petit Lob\~{a}o, P.G.S. Cardoso, S.T.R. Pinho, E.P. Borges, \emph{Braz. J. Phys.} {\bf 39}, \ 402 \\
                             \hspace*{7mm}  (2009)\\  
\noindent [30] N. Kalogeropoulos, \ \emph{Physica A} {\bf 391}, \ 1120 \ (2012)\\ 
\noindent [31] N. Kalogeropoulos, \ \emph{Physica A} {\bf 391}, \ 3435 \ (2012)\\
\noindent [32] N. Kalogeropoulos, \ \emph{Vanishing largest Lyapunov exponent and Tsallis entropy}, \\
                             \hspace*{7.5mm}  {\sf arXiv:1203.2707}\\
\noindent [33] N. Kalogeropoulos, \ \emph{Escort distributions and Tsallis entropy}, \ {\sf arXiv:1206.5127}\\ 
\noindent [34] A.J. Creaco, N. Kalogeropoulos, \ \emph{Nilpotence in Physics: the case of Tsallis entropy}, \\
                           \hspace*{7.5mm}{\sf arXiv:1209.4180}\\
\noindent [35] A. Grigor'yan, \ \emph{Heat Kernel and Analysis on Manifolds},  Studies in Adv. Math. Vol. 47, \\
                           \hspace*{6.5mm}  Amer. Math. Soc., \ Providence (2009)\\ 
\noindent [36] S. Umarov, C. Tsallis, M. Gell-Mann, S. Steinberg, \ \emph{J. Math. Phys.} {\bf 51}, \ 033502 \ (2010)\\
\noindent [37] H.J. Hilhorst, \ \emph{J. Stat. Mech.} \ P10023, \ (2010)\\
\noindent [38] M. Jauregui, C. Tsallis, \ \emph{Phys. Lett. A} {\bf 375}, \ 2085 \ (2011)\\
\noindent [39] M. Jauregui, C. Tsallis, E.M.F. Curado, \ \emph{J. Stat. Mech.} \ P10016 \ (2011)\\
\noindent [40] M.E. Taylor, \ \emph{Partial Differential Equations: Basic Theory}, \ Springer, New York (1996)\\
\noindent [41] J.H. Eschenburg, E. Heintze, \ \emph{Manuscr. Math.} {\bf 68}, \ 209 \ (1990)\\
\noindent [42] H. Karcher, \ \emph{Riemannian Comparison Constructions}, \ in \ \emph{Global Differential Geometry},\\ 
                             \hspace*{6.5mm}  S.S. Chern (Ed.), \ Math. Assoc. Amer. (1989)\\
\noindent [43] Ya. B. Zel'dovich, A.S. Kompaneets, \emph{Towards a theory of heat conduction with thermal \\
                              \hspace*{6.5mm} conductivity depending on the temparature} in \emph{Collection Dedicated to Seventieth \\ 
                              \hspace*{6.5mm} Birthday of Academician A.F. Ioffe}, \ P.I. Lukirsky (Ed.), \  Izd. Acad. Nauk SSSR, \\
                              \hspace*{6.5mm} Moscow (1950)\\ 
\noindent [44] G.I. Barenblatt, \ \emph{Prikl. Mat. Mech.} {\bf 16}, \ 67 \ (1952)\\
\noindent [45] F. Otto, \ \emph{Commun. Part. Diff. Eq.} {\bf 26}, \ 101 \ (2001)\\
\noindent [46] J. L. Vazquez, \ \emph{The Porous Medium Equation}, \ Clarendon Press, Oxford (2006)\\
\noindent [47] A. Friedman, S. Kamin, \ \emph{Trans. Amer. Math. Soc.} {\bf 262}, \ 551 \ (1980)\\
\noindent [48] M. Gromov, V.D. Milman, \ \emph{Comp. Math.} {\bf 62}, \ 263 \ (1987)\\
\noindent [49] V.D. Milman, \ \emph{Ast\'{e}risque} {\bf 157-158}, \ 273 \ (1988)\\
\noindent [50] M. Talagrand, \ \emph{Ann. Probab.} {\bf 24}, \ 1 \ (1996)\\
\noindent [51] J. Lott, C. Villani, \ \emph{Ann. Math.} {\bf 169}, \ 903 \ (2009)\\
\noindent [52] P. Erd\"{o}s, A. R\'{e}nyi, \ \emph{Publ. Math. Debrecen} {\bf 6}, \ \ 290 \ (1959)\\
\noindent [53] P. Erd\"{o}s, A. R\'{e}nyi, \ \emph{Publ. Math. Inst. Hung. Acad. Sci.} {\bf 5}, \ 17 \ (1960)\\
\noindent [54] B. Bollob\'{a}s, \ \emph{Random Graphs}, \ Academic Press, \ London (1985)\\  
\noindent [55] R. Albert, A. -L. Barab\'{a}si, \ \emph{Rev. Mod. Phys.} {\bf 74}, \ 47 \ (2002)\\ 
\noindent [56] Y. Ollivier, \ \emph{J. Funct. Anal.} {\bf 256}, \ 810 \ (2009)\\

                           \vfill

\end{document}